\documentclass[a4paper,10pt]{article}
\usepackage[dvips]{graphicx}
\usepackage{amssymb,amsmath}
\numberwithin{equation}{section}

\begin{document}

\title{ Cosmological models with non-canonical scalar and fermion  fields: k-essence, f-essence and g-essence}
\author{Ratbay Myrzakulov\footnote{Email: rmyrzakulov@gmail.com }    \\ \textit{Eurasian International Center for Theoretical Physics, Dep. Gen. $\&$ Theor.  } \\ \textit{ Phys., Eurasian National University, Astana 010008, Kazakhstan} }

\date{}

\maketitle
\begin{abstract}
In the present work, we study the cosmological model with fermionic field and with the non-canonical kinetic term (\textit{fermionic k-essence} or \textit{f-essence}). We also present some important reductions of the model   as well as its some generalizations. We also found the exact solution of the model and examine the influence of such gravity-fermion interaction on  the observed accelerated expansion of our universe. Several  fermionic and bosonic-fermionic DBI models were constructed as some particular reductions of  f-essence and g-essence, respectively and their Chaplygin gas counterparts are found. Finally, some nonlinear models of bouncing and cyclic universes are proposed.
\end{abstract}
\vspace{2cm} 

\sloppy

\section{Introduction} 
The observational evidence from different sources for the present stage of accelerated expansion of our universe has driven the quest for theoretical explanations of such feature. Assuming the validity of the theory of gravity, one attempt of explanation is the existence of an unregarded, but dominated at present time, ingredient of the energy content of the universe, known as dark energy, with unusual physical properties.  The other possibility is modifying the general theory of relativity at large scales. In cosmology, the investigation for the constituents responsible for the accelerated periods in the evolution of the universe is of great interest. The mysterious dark energy has been proposed as a cause for the late time dynamics of the current accelerated phase of the universe. 

During last years theories described by the action with non-standard kinetic terms, k-essence, attracted a considerable interest.  Such theories were first studied in the context of k-inflation \cite{Mukhanov1}, and then the k-essence models were suggested  as dynamical dark energy for solving the cosmic coincidence problem \cite{Mukhanov2}-\cite{Tsyba}.  The action of the k-essence scalar field $\phi$ minimally coupled to the gravitational field $g_{\mu\nu}$ we write in the form (see e.g. \cite{Mukhanov1}-\cite{Chiba})
\begin {equation}
S=\int d^{4}x\sqrt{-g}[R+K_1(X, \phi)],
\end{equation}
where
\begin {equation}
X=0.5g^{\mu\nu}\nabla_{\mu}\phi\nabla_{\nu}\phi,
\end{equation}
is the canonical kinetic term, $\nabla_{\mu}$ is the covariant derivative associated with metric $g_{\mu\nu}$.  The important particular reductions of the scalar k-essence (1.1) are: 

i) $K_1=A_1(X)$  (purely kinetic case), 

ii) $K_1=A_1(X)B_1(\phi)$, 

iii) $K_1=A_1(X)+B_1(\phi)$.

Note that for the FRW metric, the equations of the k-essence (1.1) have the form
\begin{eqnarray}
	3H^2-0.5\rho_k &=&0,\\ 
		2\dot{H}+3H^2+0.5p_k&=&0,\\
		K_{1X}\ddot{\phi}+(\dot{K}_{1X}+3HK_{1X})\dot{\phi}-K_{1\phi}&=&0,\\
			\dot{\rho}_k+3H(\rho_k+p_k)&=&0.
	\end{eqnarray} 
Here  the kinetic term, the energy density  and  the pressure  are given by
\begin{equation}
X=0.5\dot{\phi}^2,\quad 
\rho_k=2K_{1X}X-K_1,\quad
p_k=K_1.
\end{equation}
The system (1.3)-(1.6) was studied from different points of view. As one of example, we can consider some integrable reductions of the system (1.3)-(1.6) (see e.g. \cite{Kuralay}-\cite{Shynaray}). As an example, let us consider the case when $K_1$ has the form
\begin{equation}
K_1=F(X)-V(\phi)
\end{equation}
 and we assume that $F$ obeys one of Painlev$\acute{e}$ equations e.g. the P$_{III}$ - equation 
 \begin{equation}
F_{XX}=F^{-1}F^2_{X}-X^{-1}(F_X-\alpha F^2-\beta)+\gamma F^3+\delta F^{-1}.
\end{equation}Then the system (1.3)-(1.6) takes the form
\begin{eqnarray}
	3H^2-0.5\rho_k &=&0,\\ 
		2\dot{H}+3H^2+0.5p_k&=&0,\\
		F_{X}\ddot{\phi}+(\dot{F}_{X}+3HF_{X})\dot{\phi}+V_{\phi}&=&0,\\
		F_{XX}=F^{-1}F^2_{X}-X^{-1}(F_X-\alpha F^2-\beta)+\gamma F^3+\delta F^{-1}&=&0,\\
			\dot{\rho}_k+3H(\rho_k+p_k)&=&0.
	\end{eqnarray} 
	Another class integrable reductions of the system (1.3)-(1.6) we can construct e.g. if we demand that $\phi$ satisfies one of Painlev$\acute{e}$ equations e.g. the P$_{I}$ - equation. Then the system (1.3)-(1.6) becomes
	\begin{eqnarray}
	3H^2-0.5\rho_k &=&0,\\ 
		2\dot{H}+3H^2+0.5p_k&=&0,\\
		K_{1X}\ddot{\phi}+(\dot{K}_{1X}+3HK_{1X})\dot{\phi}-K_{1\phi}&=&0,\\
			\phi_{HH}-6\phi^2-H&=&0,\\
			\dot{\rho}_k+3H(\rho_k+p_k)&=&0.
	\end{eqnarray} 
 In the recent years several approaches were made to explain the accelerated expansion by choosing fermionic fields as the gravitational sources of energy (see e.g. refs. \cite{Ribas1}-\cite{Armendariz-Picon}). In particular, it was shown that the fermionic  field plays very important role in: i) isotropization of initially anisotropic spacetime; ii) formation of singularity free cosmological solutions; iii) explaining late-time acceleration. In the present work, we study the cosmological model with fermionic field, the  M$_{33}$ - model,  which has the non-canonical kinetic term (\textit{ f-essence}). We examine the influence of  such gravity-fermionic interaction on   the accelerated expansion of the Universe.
The formulation of the gravity-fermionic theory has been discussed in detail elsewhere \cite{Weinberg}-\cite{Birrell}., so we will only present the result here.

\section{Einstein-Dirac equations}
In order to have this work self-consistent, in this section we present briefly the techniques that are used to include fermionic sources in the Einstein theory of gravitation and for a more detailed analysis the reader is referred to \cite{Weinberg}-\cite{Birrell}.   The general covariance principle imposes that the Dirac-Pauli matrices $\gamma^a$ must be replaced by their generalized counterparts $\Gamma^{\mu}=e^{\mu}_{a}\gamma^{a}$, whereas the generalized Dirac-Pauli matrices satisfy the extended Clifford algebra, i.e., $\{\Gamma^{\mu}, \Gamma^{\nu}\}=2g^{\mu\nu}$. 	   The Einstein-Dirac action reads as 
\begin {equation}
S=\int d^{4}x\sqrt{-g}[R+\epsilon Y-V],
\end{equation} 
 where $\epsilon=\pm 1$ ($\epsilon=1$ is the usual case and $\epsilon=-1$ is the phantom case) and 
 \begin{equation}
Y=0.5i[\bar{\psi}\Gamma^{\mu}D_{\mu}\psi-(D_{\mu}\bar{\psi})\Gamma^{\mu}\psi],\quad V=V(\bar{\psi},\psi).
\end{equation}  Here $\psi$ and $\bar{\psi}=\psi^+\gamma^0$ denote the fermionic field and its adjoint, respectively and $R$ is the Ricci scalar. The  covariant derivatives are given by 
\begin{equation}
D_{\mu}\psi=\partial_{\mu}\psi-\Omega_{\mu}\psi, \quad D_{\mu}\bar{\psi}=\partial_{\mu}\bar{\psi}+\bar{\psi}\Omega_{\mu},
\end{equation}
where the spin connection $\Omega_{\mu}$ is given by
 \begin{equation}
\Omega_{\mu}=-0.25g_{\mu\nu}[\Gamma^{\nu}_{\sigma\lambda}-e^{\nu}_{b}(\partial_{\sigma}e^{b}_{\lambda})]\gamma^{\sigma}\gamma^{\lambda},
\end{equation}
with $\Gamma^{\nu}_{\sigma\lambda}$ denoting the Christoffel symbols.
The closed system of the equations for the model (2.1) looks like 
 \begin{eqnarray}
R_{\mu\nu}-0.5Rg_{\mu\nu}+T_{\mu\nu}&=&0,\\
	i\Gamma^{\mu}D_{\mu}\psi-\frac{dV}{d\bar{\psi}}&=&0,\\ 
		iD_{\mu}\bar{\psi}\Gamma^{\mu}+\frac{dV}{d\psi}&=&0,\\
	\dot{\rho}_f+3H(\rho_f+p_f)&=&0,
	\end{eqnarray}
	where the density of energy and pressure are given by
		\begin{equation}
\rho_f=V,\quad p_f=\epsilon Y-V.
\end{equation}
\section{F-essence}

Let us now we consider the  M$_{33}$ - model or \textit{f-essence}, which has  the  action
\begin {equation}
S=\int d^{4}x\sqrt{-g}[R+K_2(Y, \psi, \bar{\psi})],
\end{equation} 
 where $K_2$ is some function of its arguments and the canonical kinetic term has the form\begin{equation}
Y=0.5i[\bar{\psi}\Gamma^{\mu}D_{\mu}\psi-(D_{\mu}\bar{\psi})\Gamma^{\mu}\psi].
\end{equation}
	We work with a space-time metric of the form 
	\begin{equation}
ds^2=-dt^2+a^2(dx^2+dy^2+dz^2),
\end{equation}
that is the FRW metric.  
For this metric,the vierbein is chosen to be
	\begin{equation}
	(e_a^\mu)=diag(1,1/a,1/a,1/a),\quad 
(e^a_\mu)=diag(1,a,a,a).
\end{equation} The Dirac matrices of curved spacetime $\Gamma^\mu$ are
\begin{equation}
\Gamma^0=\gamma^0, \quad \Gamma^1=a^{-1}\gamma^1, \quad \Gamma^2=a^{-1}\gamma^2, \quad \Gamma^3=a^{-1}\gamma^3, 
\end{equation}
\begin{equation}
\Gamma_0=\gamma^0, \quad \Gamma_1=a\gamma^1, \quad \Gamma_2=a\gamma_2, \quad \Gamma_3=a\gamma_3.
\end{equation}Hence we get
\begin{equation}
\Omega_0=0, \quad \Omega_1=0.5\dot{a}\gamma^1\gamma^0, \quad \Omega_2=0.5\dot{a}\gamma^2\gamma^0, \quad \Omega_3=0.5\dot{a}\gamma^3\gamma^0. 
\end{equation}
 We now ready to write  the equations of the M$_{33}$ - model (3.1). We have 
	\begin{eqnarray}
	3H^2+0.5[K_{2}+0.5(K_{2\psi}\psi+K_{2\bar{\psi}}\bar{\psi})]&=&0,\\ 
		2\dot{H}+3H^2+0.5K_{2}&=&0,\\
		K_{2Y}\dot{\psi}+0.5(3HK_{2Y}+\dot{K}_{2Y})\psi-i\gamma^0K_{2\bar{\psi}}&=&0,\\ 
K_{2Y}\dot{\bar{\psi}}+0.5(3HK_{2Y}+\dot{K}_{2Y})\bar{\psi}+iK_{2\psi}\gamma^{0}&=&0,\\
	\dot{\rho}_f+3H(\rho_f+p_f)&=&0,
	\end{eqnarray} 
where $Y=0.5i(\bar{\psi}\gamma^{0}\dot{\psi}-\dot{\bar{\psi}}\gamma^{0}\psi)$ and 
\begin{equation}
\rho_f=K_{2Y}Y-K_2=-[K_{2}+0.5(K_{2\psi}\psi+K_{2\bar{\psi}}\bar{\psi})],\quad
p_f=K_2
\end{equation}
are the energy density and pressure of the fermionic field.
	If $K_2=Y-V$, then from the system (3.8)-(3.12) we get the corresponding equations of   the Einstein-Dirac model. 
\subsection{Submodels}
\subsubsection{The M$_{33A}$ - model}
	The Langrangian of the M$_{33A}$ - model has the form
	\begin{equation}
K_2=A_{2}(Y).
\end{equation}
So the M$_{33A}$ - model is the purely kinetic fermionic k-essence.
\subsubsection{The M$_{33B}$ - model}
		The Langrangian of the M$_{33B}$ - model reads as
	\begin{equation}
K_2=A_{2}(Y)B_2(\psi, \bar{\psi}).
\end{equation}
\subsubsection{The M$_{33C}$ - model}
		The Langrangian of the M$_{33A}$ - model has the form
	\begin{equation}
K_2=A_{2}(Y)+B_2(\psi, \bar{\psi}).
\end{equation}
\subsection{Solution}
In this subsection we want to construct a solution of the M$_{33}$-model. Let $K_2=K_2(Y,u)$, where $u=\bar{\psi}\psi$. Then the system (3.8)-(3.12) becomes
	\begin{eqnarray}
	3H^2+0.5[K_{2}+K_{2}^{'}u]&=&0,\\ 
		2\dot{H}+3H^2+0.5K_{2}&=&0,\\
		K_{2Y}\dot{\psi}+0.5(3HK_{2Y}+\dot{K}_{2Y})\psi-i\gamma^0K^{'}_{2}\psi&=&0,\\ 
K_{2Y}\dot{\bar{\psi}}+0.5(3HK_{2Y}+\dot{K}_{2Y})\bar{\psi}+iK^{'}_{2}\bar{\psi}\gamma^{0}&=&0,\\
	\dot{\rho}_f+3H(\rho_f+p_f)&=&0,
	\end{eqnarray}
where $K^{'}_{2}=dK_{2}/du$ and
\begin{equation}
\rho_f=-[K_{2}+K^{'}_{2}u].\quad
p_f=K_2.
\end{equation}
We now consider the submodel (3.16) that is the  M$_{33C}$-model, where we assume that $A_2=\alpha Y^n, \quad B_2=\beta u^m$ that is the case $K_2=\alpha Y^n+ \beta u^m.$ Let $a=a_0t^{\lambda}$. Then we have the following solution
\begin {equation}
Y=\left\{\left[-\frac{6m\lambda^2-4(m+1)\lambda}{\alpha m}\right]t^{-2}\right\}^{\frac{1}{n}},\quad 
u=\left\{\left[\frac{-4\lambda}{\beta  m}\right]t^{-2}\right\}^{\frac{1}{m}},
\end{equation}
where
\begin {equation}
\lambda=\frac{2n-2m+2mn}{3mn}, \quad a_{0}=\sqrt[3]{\frac{c}{\alpha n}\left[-\frac{6m\lambda^2-4(m+1)\lambda}{\alpha m}\right]^{\frac{1-n}{n}}\left[-\frac{4\lambda}{\beta m}\right]^{-\frac{1}{m}}}.\end{equation}
Finally we present the following formulas
\begin {equation}
u=ca^{-3}K_{2Y}^{-1}, \quad \psi_j=c_ja^{-1.5}K_{2Y}^{-0.5}e^{i\gamma^{0}\int K^{'}_{2}K_{2Y}^{-1}dt},
\end{equation}
where $c=|c|^{2}_1+|c|^{2}_2-|c|^{2}_3-|c|^{2}_4, \quad c_j=consts.$ 
\subsection{f-DBI }
Let us rewrite the action (3.1) as 
 \begin {equation}
S=\int d^{4}x\sqrt{-g}[R+2K_2(Y, \psi, \bar{\psi})].
\end{equation} 
 For the FRW metric the corresponding system of equations takes the form
 \begin{eqnarray}
	3H^2-\rho_f&=&0,\\ 
		2\dot{H}+3H^2+p_f&=&0,\\
		K_{2Y}\dot{\psi}+0.5(3HK_{2Y}+\dot{K}_{2Y})\psi-i\gamma^0K_{2\bar{\psi}}&=&0,\\ 
K_{2Y}\dot{\bar{\psi}}+0.5(3HK_{2Y}+\dot{K}_{2Y})\bar{\psi}+iK_{2\psi}\gamma^{0}&=&0,\\
	\dot{\rho}_f+3H(\rho_f+p_f)&=&0,
	\end{eqnarray} 
where $Y=0.5i(\bar{\psi}\gamma^{0}\dot{\psi}-\dot{\bar{\psi}}\gamma^{0}\psi)$ and 
\begin{equation}
\rho_f=YK_{2Y}-K_2,\quad
p_f=K_2. 
\end{equation}
 Now we want present some examples fermionic DBI models or short f-DBI.
 
 \subsubsection{Tachyonic models}
 i) First let us consider the following tachyonic models 
 \begin{equation}
K_2=-U\sqrt{1-\alpha Y}, 
\end{equation}where
\begin{equation}
U=U(\bar{T}, T), \quad Y=0.5i(\bar{T}\gamma^{0}\dot{T}-\dot{\bar{T}}\gamma^{0}T) 
\end{equation}
and $T$ is the fermionic tachyon field. Then we get the following EoS
\begin{equation}
p=-\rho\mp\sqrt{\rho^2-U^2}, 
\end{equation}
where 
\begin{equation}
\rho=\frac{0.5U}{\sqrt{1-\alpha Y}}+0.5U\sqrt{1-\alpha Y}. 
\end{equation}
ii) Let the f-essence Lagrangian has the form 
 \begin{equation}
K_2=-U\sqrt{1-\alpha Y^2}, 
\end{equation}where
\begin{equation}
U=U(\bar{T}, T), \quad Y=0.5i(\bar{T}\gamma^{0}\dot{T}-\dot{\bar{T}}\gamma^{0}T) 
\end{equation}
and $T$ is the fermionic tachyon field. Then we get the following EoS
\begin{equation}
p=-\frac{U^2}{\rho}, 
\end{equation}
where 
\begin{equation}
\rho=\frac{U}{\sqrt{1-\alpha Y^2}}. 
\end{equation}
So this model corresponds to the Chaplygin gas \cite{Kamenshchik}.
\subsubsection{Dark energy models}
a) Now we want  consider the following DBI model 
 \begin{equation}
K_2=\epsilon U(\sqrt{1-\alpha \frac{Y}{U}}-1)+V, 
\end{equation}where
\begin{equation}
U=U(\bar{\psi}, \psi), \quad Y=0.5i(\bar{\psi}\gamma^{0}\dot{\psi}-\dot{\bar{\psi}}\gamma^{0}\psi), \quad V=V(\bar{\psi}, \psi) 
\end{equation}
and $\psi$ is the classical commuting fermionic  field. Then we get the following EoS
\begin{equation}
p=-2\bar{\rho}\pm\sqrt{4\bar{\rho}^2-2\epsilon^2U^2}+V-\epsilon U, 
\end{equation}
where $\bar{\rho}=\rho-\epsilon U+V$ and the energy density is given by
\begin{equation}
\rho=-0.5\epsilon U\sqrt{1-\alpha \frac{Y}{U}}-\frac{0.5\epsilon U}{\sqrt{1-\alpha \frac{Y}{U}}}+\epsilon U-V. 
\end{equation}
b) Our second example reads as
\begin{equation}
K_2=\epsilon U(\sqrt{1-\alpha \frac{Y^2}{U}}-1)+V, 
\end{equation}where
\begin{equation}
U=U(\bar{\psi}, \psi), \quad Y=0.5i(\bar{\psi}\gamma^{0}\dot{\psi}-\dot{\bar{\psi}}\gamma^{0}\psi), \quad V=V(\bar{\psi}, \psi) 
\end{equation}
and $\psi$ is the classical commuting fermionic  field. Then we get the following EoS
\begin{equation}
p=-\frac{\epsilon^2U^2}{\rho-\epsilon U+V}-\epsilon U+V, 
\end{equation}
where the energy density is given by
\begin{equation}
\rho=-\frac{\epsilon U}{\sqrt{1-\alpha \frac{Y^2}{U}}}+\epsilon U-V. 
\end{equation} From (3.47) follows that as $V=\epsilon U$ this model becomes
\begin{equation}
p=-\frac{\epsilon^2U^2}{\rho}
\end{equation}
that corresponds to the Chaplygin gas \cite{Kamenshchik}.
	\subsection{Integrable f-essence models}
	One of interesting class of f-essence models is integrable models (see e.g. \cite{Kuralay}-\cite{Shynaray}). Let the Lagrangian of f-essence has the form
		\begin{equation}
K_{2}=F(Y)-V(\bar{\psi}, \psi).\end{equation}
	 As an example, here we consider the following particular f-essence model when $F(Y)$  obeys the equation
	\begin{equation}
F_{YY}=2F^3+YF+\alpha,\end{equation}
where $\alpha$ is some constant. It is nothing but the P$_{II}$ - equation so that is integrable. Now the equations of f-essence (3.27)-(3.31)  take the form
\begin{eqnarray}
	3H^2-\rho_f&=&0,\\ 
		2\dot{H}+3H^2+p_f&=&0,\\
		F_{Y}\dot{\psi}+0.5(3HF_{Y}+\dot{F}_{Y})\psi+i\gamma^0V_{\bar{\psi}}&=&0,\\ 
F_{Y}\dot{\bar{\psi}}+0.5(3HF_{Y}+\dot{F}_{Y})\bar{\psi}-iV_{\psi}\gamma^{0}&=&0,\\
F_{YY}-2F^3-YF-\alpha&=&0,\\
	\dot{\rho}_f+3H(\rho_f+p_f)&=&0.
	\end{eqnarray} 
To construct solutions of this system we start from the equation (3.56). The equation 
(3.56) has the following particular solutions (see e.g. \cite{Kuralay}-\cite{Shynaray})
\begin{eqnarray}
  F&\equiv &F(Y;1.5)=\psi-(2\psi^2+Y)^{-1},\\
 F&\equiv &F(Y;1)=-\frac{1}{Y}, \\ 
  F&\equiv &F(Y;2)=\frac{1}{Y}-\frac{3Y^2}{Y^3+4},\\
  F&\equiv& F(Y;3)=\frac{3Y^2}{Y^3+4}-\frac{6Y^2(Y^3+10)}{Y^6+20Y^3-80},\\
F&\equiv& F(Y;4)=-\frac{1}{Y}+\frac{6Y^2(Y^3+10)}{Y^6+20Y^3-80}-\frac{9Y^5(Y^3+40)}{Y^9+60Y^6+11200},\\
 F&\equiv& F(Y;0.5\epsilon)=-\epsilon\psi
 \end{eqnarray}
and so on. Here 
 \begin{equation}
\psi=(\ln{\phi})_{Y}, \quad  \phi(Y)=C_1Ai(-2^{-1/3}Y)+C_2Bi{(-2^{-1/3}Y)},\end{equation} 
where
 $C_i=consts$ and $Ai(x), Bi(x)$ are Airy functions.

	\section{G-essence}
\subsection{Equations}
We now would like to present the M$_{34}$ - model  (the generalized k-essence or \textit{g-essence} for short) which has the following action 
\begin {equation}
S=\int d^{4}x\sqrt{-g}[R+2K(X, Y, \phi, \psi, \bar{\psi})].
\end{equation}For the FRW metric (3.3), the  equations of \textit{g-essence} (4.1) have the form
\begin{eqnarray}
	3H^2-\rho &=&0,\\ 
		2\dot{H}+3H^2+p&=&0,\\
		K_{X}\ddot{\phi}+(\dot{K}_{X}+3HK_{X})\dot{\phi}-K_{\phi}&=&0,\\
		K_{Y}\dot{\psi}+0.5(3HK_{Y}+\dot{K}_{Y})\psi-i\gamma^0K_{\bar{\psi}}&=&0,\\ 
K_{Y}\dot{\bar{\psi}}+0.5(3HK_{Y}+\dot{K}_{Y})\bar{\psi}+iK_{\psi}\gamma^{0}&=&0,\\
	\dot{\rho}+3H(\rho+p)&=&0.
	\end{eqnarray} 
Here  the kinetic terms, the energy density  and  the pressure  are given by
\begin{equation}
X=0.5\dot{\phi}^2,\quad  Y=0.5i(\bar{\psi}\gamma^{0}\dot{\psi}-\dot{\bar{\psi}}\gamma^{0}\psi)
  \end{equation}
  and 
\begin{equation}
\rho=2XK_{X}+YK_{Y}-K,\quad
p=K.
\end{equation}
Note that the model (4.1)  contents some important particular submodels. For example:

i) the scalar  k-essence (1.1) as $K=K_1(X, \phi)$;

ii) the M$_{33}$ - model  (3.1) as $K=K_2(Y,  \psi, \bar{\psi})$;

iii) the M$_{34A}$ - model   as $K=K_1(X, \phi)K_2(Y,  \psi, \bar{\psi})$;

vi) the M$_{34B}$ - model   as $K=K_1(X, \phi)+K_2(Y,  \psi, \bar{\psi})$.

\subsection{Bosonic-fermionic Chaplygin gas}
As for k-essence and f-essence cases we can find the bosonic-fermionic Chaplygin gas from g-essence. To do it, let us consider the following particular g-essence model
\begin{equation}
K=V_0[1+V_1X^{1/2n}+V_2Y^{1/n}+V_3X^{\nu}Y^{(1-2n\nu)/n}]^n,\end{equation}
where $V_j=V_j(\phi, \bar{\psi}, \psi)$ and $n,\nu=consts$. Then we have
\begin{eqnarray}
	p&=&V_0[1+V_1X^{1/2n}+V_2Y^{1/n}+V_3X^{\nu}Y^{(1-2n\nu)/n}]^n,\\ 
	\rho&=&-V_0[1+V_1X^{1/2n}+V_2Y^{1/n}+V_3X^{\nu}Y^{(1-2n\nu)/n}]^{n-1}.
	\end{eqnarray} Hence we get the generalized bosonic-fermionic Chaplygin gas
	\begin{equation}
p=-\frac{(-V_0)^{\frac{1}{1-n}}}{\rho^{\frac{n}{1-n}}}=-\frac{A}{\rho^{\alpha}},\end{equation}
	where $\alpha=n/(1-n),\quad A=(-V_0)^{\frac{1}{1-n}}$.  The case $n=0.5$ corresponds to the  bosonic-fermionic Chaplygin gas.
	\subsection{Bosonic-fermionic DBI}
Consider the following bosonic-fermionic DBI model
\begin{equation}
K=V_0[1+V_1X^{1/2n}+V_2Y^{1/n}+V_3X^{\nu}Y^{(1-2n\nu)/n}]^n+V_4,\end{equation}
where $V_j=V_j(\phi, \bar{\psi}, \psi)$ and $n,\nu=consts$. If $n=0.5$ then it takes the form
\begin{equation}
K=V_0\sqrt{1+V_1X+V_2Y^{2}+V_3X^{\nu}Y^{2(1-\nu)}}+V_4.\end{equation}
 From (4.14)  we have
\begin{eqnarray}
	p&=&V_0[1+V_1X^{1/2n}+V_2Y^{1/n}+V_3X^{\nu}Y^{(1-2n\nu)/n}]^n+V_4,\\ 
	\rho&=&-V_0[1+V_1X^{1/2n}+V_2Y^{1/n}+V_3X^{\nu}Y^{(1-2n\nu)/n}]^{n-1}-V_4.
	\end{eqnarray} Hence we get the generalized bosonic-fermionic Chaplygin gas
	\begin{equation}
p=-\frac{(-V_0)^{\frac{1}{1-n}}}{(\rho+V_4)^{\frac{n}{1-n}}}=-\frac{A}{(\rho+V_4)^{\alpha}},\end{equation}
	where $\alpha=n/(1-n),\quad A=(-V_0)^{\frac{1}{1-n}}$.  The case $n=0.5$ corresponds to the  bosonic-fermionic Chaplygin gas. The interesting particular case of the model (4.14) we get if we set
	\begin{equation}
K=V_0\{[1+V_1^{'}V_0^{-1}X^{1/2n}+V^{'}_2V_0^{'}Y^{1/n}+V^{'}_3V_0^{'}X^{\nu}Y^{(1-2n\nu)/n}]^n-1\}+V(\phi, \bar{\psi}, \psi),\end{equation}
where $V_j^{'}=	V_j^{'}(\phi,\bar{\psi},\psi)$. Let $n=0.5$. Then the bosonic-fermionic DBI (4.18) becomes
	\begin{equation}
K=V_0[\sqrt{1+\beta_1 X+\beta_2Y^{2}+\beta_3X^{\nu}Y^{2(1-\nu)}}-1]+V(\phi, \bar{\psi}, \psi),\end{equation}
where $\beta_j=V_j^{'}V_0^{-1}$. 
\section{On some models of bouncing and cyclic universes}	
Finally we would like to note that it is interesting to find some f-essence  models which describe the cyclic and bouncing universes. Here we briefly  present some of such models (see e.g. \cite{Kuralay}-\cite{Shynaray}).

i) Consider the following model
\begin {equation}
a_t-a_{\xi\xi t}+2\kappa a_{\xi}+3aa_{\xi}-2a_{\xi}a_{\xi\xi}-aa_{\xi\xi\xi}=0\end{equation}
or its twin
\begin {equation}
a_\xi-a_{tt\xi}+2\kappa a_{t}+3aa_{t}-2a_{t}a_{tt}-aa_{ttt}=0\end{equation}
which is nothing but the Camassa-Holm equation. Here $a=a(t,\xi)$ is the scale factor, $\kappa=const$ and $\xi$ is some artificial coordinate e.g. $\xi=\Lambda$ (cosmological constant), $\xi=G$ (Newton's gravitational constant), $\xi=k$  (spatial curvature) or $\xi=c$ (speed of light) see e.g. 
\begin {equation}
\frac{\dot{a}}{a}^2=\frac{8\pi G}{3}\rho-\frac{kc^2}{a^2}+\frac{\Lambda c^2}{3},\quad \frac{\ddot{a}}{a}=-\frac{4\pi G}{3}(\rho+3c^{-2}p)+\frac{\Lambda c^2}{3}.\end{equation}
In the case of no linear dispersion, $\kappa=0$ and $\xi=\Lambda$, the equations (5.1) and (5.2) possess singular solutions in the form of peaked travelling waves called peakons
\begin {equation}
a=0.5\eta e^{-|\xi-\eta t|}\end{equation}
	and \begin {equation}
a=0.5\eta e^{-|t-\eta \xi|},\end{equation}
	respectively. They are  1-peakon solutions. The multi-peakon solution versions of these solutions have the form
	\begin {equation}
a=\sum_{j=1}^{n}r_{j}(t) e^{-2|\xi-q_j(t)|}\end{equation}
	and \begin {equation}
a=\sum_{j=1}^{n}r_{j}(\xi) e^{-2|t-q_j(\xi)|},\end{equation}
respectively.	Our idea is that the n-peakon solutions can describe the cyclic and bounce universes. Hope that we will return to this question in the future from the more detail position.

 	ii) Our second  example is  the following model  [here we set e.g. $\xi=\Lambda$ and $a_{\Lambda}=da/d\Lambda, \quad a_t=da/dt$ etc]:
 	\begin{equation}
a_t=0.75(a^2)_{\Lambda}
 \end{equation}
 or its twin
 	\begin{equation}
a_{\Lambda}=0.75(a^2)_{t}
 \end{equation}
 which are known to develop shocks. Eqs. (5.8) and (5.9)  are nothing but the dispersionless Korteweg-de Vries equations (dKdVE)  	 or Riemann equations. It is well-known that the equations (5.8) and (5.9) have the following solutions 
 \begin{equation}
a(\Lambda, t)=h(\Lambda-0.75a t)
 \end{equation}
 and\begin{equation}
a(t, \Lambda)=h(t- 0.75a\Lambda),
 \end{equation}	 	
 respectively. Here $h$ is an arbitrary function. Also we see e.g. from the solution (5.10) that the velocity of a point of the wave, with constant amplitude $a$, is proportional to its amplitude leading to the "breaking" of the wave. Also we note that the wave also develops  discontinuities in its evolution. Let's now we give the following particular solutions of the equations (5.8) and (5.9):
  \begin{equation}
a=(\beta_1 +\beta_2\Lambda)(\beta_3-1.5\beta_2 t)^{-1}
 \end{equation}
 and\begin{equation}
a=(\beta_1 +\beta_2 t)(\beta_3-1.5\beta_2 \Lambda)^{-1},
 \end{equation}	 	
 respectively, where $\beta_i=consts$.

	\section{Conclusion}We briefly summarize the present work. We first derived the equations of the   M$_{33}$ - model for the FRW space-time.  Then we found their exact solution for the \textit{ f-essence} $K_2=\alpha Y^n+ \beta u^m.$ Finally,   let us we present  the expressions for the equation of state and deceleration parameters. For the our particular solution  (3.19)  they take the form
	\begin {equation}
w_f=-1+\frac{mn}{n-m+mn}, \quad q=-1+\frac{3mn}{2(n-m+mn)}.\end{equation}
These formulas  tell us   that the M$_{33}$ - model  can describes the observed accelerated expansion of our universe.  Also we have been constructed the fermionic DBI models from f-essence.  Finally, some nonlinear models of bouncing and cyclic universes are proposed.

\end{document}